\documentclass{article}
\usepackage{epsf}
\usepackage{emulateapj}
\input{psfig}
\begin{document}

\title{ Early Chandra X-ray Observations of Eta Carinae} 

\author{F. D. Seward, Y. M. Butt, M. Karovska, A. Prestwich. E. M. Schlegel}
\affil{Smithsonian Astrophysical Observatory, 60 Garden St., Cambridge, MA 02138}

\author{M. Corcoran}
\affil{Laboratory for High Energy Astrophysics, Goddard Space Flight Center, Greenbelt, MD 20771}

\begin{abstract}  

Sub-arcsecond resolution Chandra observations of $\eta$ Car reveal a
$40^{\prime\prime} \times 70^{\prime\prime}$ 
ring or partial shell of X-ray emission surrounding an
unresolved, bright, central source.  The spectrum of the central source
is strongly absorbed and can be fit with a high-temperature thermal
continuum and emission lines.  The surrounding shell is well outside
the optical/IR bipolar nebula and is coincident with
the Outer Shell of $\eta$ Car.
The X-ray spectrum of the Shell is much softer than
that of the central source.  The X-ray Shell is irregular and only
correlates well with optical features where a bright X-ray knot
coincides with a bright feature of the Outer Shell.  Implications for
the binary model of the central source are discussed.

\end{abstract}




%

\section{Introduction}

Eta Carinae has long been an object of mystery, a unique
optical variable (see review by Davidson \& Humphreys, 1997). 
Initially of 4th to 2nd magnitude, it brightened to
first magnitude in the 1830s and reached a maximum brightness of -1 in
1843. Between 1857 and 1869 it declined steadily from 1st to 7th
magnitude.  It underwent a second outburst in 1890 (Humphreys,
Davidson, \& Smith, 1999), 
declined to and remained at 8th magnitude until about 1940, and
has since brightened to almost 5th magnitude (Davidson et al 1999).
The central source of Eta Carinae has long been obscured by a small
optical nebula, the homunculus, which HST observations have revealed
to be a striking bipolar nebula (Hester et al, 1991; Humphreys \& Davidson 1994).
Measured expansion of the lobes indicate an origin at the time of the
``Great Eruption'' in 1843 (Walborn et al, 1978).  There is also a
rapidly expanding disk in the plane between the bipolar lobes which
seems to have originated in the later eruption in about 1890 (Smith \&
Gehrz, 1998).  A faint
"Outer Shell", with knots of fast-moving material, surrounds the
bipolar structure (Walborn 1976). \\ 

Eta Carinae is now the brightest extra-solar infrared source at a
wavelength of 20 microns.  We will use a distance of 2.3 kpc in
this paper (Davidson \& Humphreys, 1997) but note that most of the
previous X-ray work assumed a slightly larger distance of 2.6 kpc.
An infrared luminosity of $5 \times 10^6$ L$_{\odot}$ 
was derived by Westphal \& Neugebauer (1969) which is only a factor of
$\approx$4 less than the optical
luminosity during the Great Eruption (Davidson \& Humphreys, 1997).
It is now believed that the source of energy is a massive
object, or objects, at the center and that surrounding dust then re-radiates
the energy in the infrared.  Indeed, infrared emission is observed
to come from the bipolar nebula and the equatorial disk (Smith et al,
1998). The central source, however, remains obscure at optical and
infrared wavelengths. \\

Since X-rays can penetrate dust, the X-ray band holds the promise of 
revealing more of the central object.
The first convincing observation was with Einstein
(Chlebowski et al 1984).  This showed emission from an unresolved
source in the vicinity of the central object and from an 
irregular outer region.  This
structure was confirmed with ROSAT observations by Corcoran et al
(1995) who discovered that the central source was variable.  The
Einstein spectrum was interpreted as the sum of a
hard, strongly-absorbed component and a soft component.  A later
observation with ASCA (Corcoran et al, 1998) showed the same dual
nature and, with the better ASCA energy resolution, emission lines
from Mg, Si, S and Fe 
were detected in the spectrum.  Neither the Einstein IPC nor the ASCA
SIS instruments could spatially resolve the central source from the
surrounding shell.  A campaign to monitor high energy X-rays with XTE
has been in progress since April 1996.   The variability observed is
remarkable, showing a possible 85 day periodicity and a 3-month
X-ray minimum (Ishibashi et al, 1999).  These results have been interpreted
with a model invoking colliding stellar winds and a massive binary
system (Corcoran et al, 2000b). \\

\section{Chandra Observations}

$\eta$ Car was one of the first science observations done with Chandra
(Weisskopf, O'Dell, \& VanSpeybroeck 1996).  
To demonstrate the capability
of the observatory, Eta Carinae was observed on 7 September 1999 with the
Chandra ACIS-I detector, first in TE (timed exposure) mode for 10,800 s,
then in CC (continuous clocking) mode for 12,200 s.  During this
period (and unbeknownst at the time) the energy resolution of the
detector was degrading significantly as a result of radiation
damage.  Although the energy resolution
of these spectra is considerably better than the present capability of
the instrument, the calibration for this period of rapidly-changing
characteristics is poor. \\

The TE-mode image is shown in Figure 1.	
A partial ring or shell of
soft emission with dimension $\sim 40^{\prime\prime}\times 
70^{\prime\prime}$ surrounds a bright central
source. It is 
open in the south and lies closer to the central source in the west than in
the east.  There are two bright knots.  Emission from the ring is soft
whereas the spectrum of the central source is hard and strongly
absorbed.  At energies below 1.0 keV, only the ring is seen. \\
 
The geometry and weakness of interior diffuse emission
indicate a geometry more ring-like than shell-like.  The orientation of
this material, however,
does not have a sensible relationship to the geometry of the bipolar
nebula.  The ring, e.g., is not in the equatorial plane where it might
first be expected.  The most likely possibility is that we are
seeing clumps of material in an irregular shell so we will refer to
this material as the (Outer) Shell, but will use a partial torus to
calculate the density. \\

\begin{center}
\psfig{figure=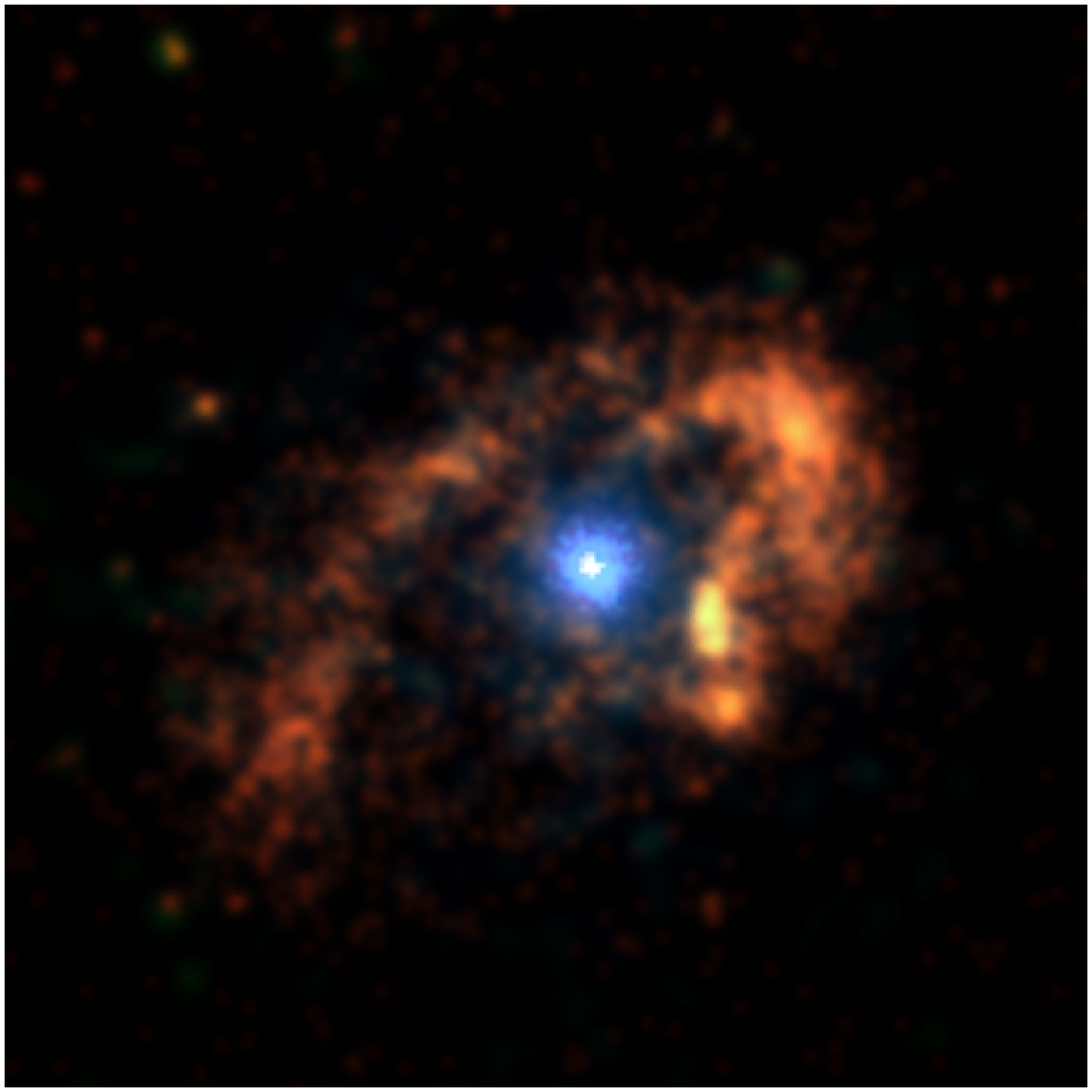,width=3.5in}
\begin{minipage}[h]{3.5in}
{\small
Figure 1:Contours of 2.1-8.0 keV surface brightness overlaid on a 
6590$\AA$ HST image of $\eta$ Car printed to show the Outer Shell.
Contour spacing is proportional to the square root of the surface 
brightness.  The bright X-ray source is coincident with the central
source of the bipolar nebula.  Closely-spaced contours around
this source indicate scattering from the telescope mirror but
the faintest contour partially follows the the Outer Shell.  
} 
\end{minipage}
\end{center}

Figure 1 was made using an adaptive smoothing algorithm (CSMOOTH) to
post-process the $\eta$ Car images.  This smooths a two-dimensional
image in such a way that the smoothing scale is increased until the
total number of counts under the kernel exceeds a value that is
determined from a preset significance and the expected number of
background counts in the kernel area. The scale is determined by
convolving the raw image with a Gaussian kernel whose size grows from
a small initial value to the maximal allowable value corresponding to
the size of the raw image itself. \\ 

Data were analyzed using the Chandra CIAO software and the FTOOLS
XSPEC.  Both TE and CC mode data were first cleaned by rejecting ACIS
event-grades 1,5, and 7  (`Chandra Proposers' Observatory Guide,
2000).  Since $\eta$ Car is a bright source we did
not do any background filtering.  The ACIS focal-plane temperature 
at the time of
the observation was $-100^{\circ}$ C.  Since no calibration has yet
been done for this temperature, we used spectral response
matrices for $-90^{\circ}$ and shifted the
gain so the model spectrum reproduced the observed Ir (the mirror
coating) absorption edge at 2.1 keV.  The observed spectral structure
indicates that the resolution is close to that observed in the
$-90^{\circ}$ calibrations.  the gain shift was also checked against
a better-known Cas A spectrum taken a few days earlier.  
The lack of calibrated response matrices,
however, prevents an accurate spectral analysis, particularly at low
energies. \\ 

\section{Comparison with Bipolar Nebula and Outer Shell}

Figure 2 shows 0.1-0.8 keV X-ray	
emission superimposed on the HST WFPC image 
(courtesy Jon Morse) with the F658N filter.  This
filter is centered at 6590 \AA, and passes [NII] lambda 6583 and scattered
H-alpha (Morse et al., 1998).   X-ray emission
from the central source is strongly absorbed, so only the 
extended soft emission from the Shell is visible.  Note that
there are no X-ray features corresponding with the bright
optical bipolar nebula.  The extended X-ray Shell is, instead, clearly
associated with compact knots of line emission described by Morse et al
(1998) as being in the ``Outer Debris Field'', which is the same region as
Walborn's ``Outer Shell''.  The brightest X-ray
knot coincides exactly with a bright optical patch in the SW Outer
Shell (Walborn's ``S condensation'').  The X-ray
maximum which forms the NW part of the Shell is at the outer boundary
of the Debris Field (the ``W arc'') and the maximum at the 
SE extreme of the X-ray
Shell coincides with the SE extension of the Debris Field 
(the ``E condensations''). \\

Figure 3 shows a higher energy band; here   
2.1-8.0 keV X-ray brightness contours are superimposed on the same F658N image.
The central source now dominates the X-ray image but there is also a hint
of emission from the Outer Shell. This will be discussed in Section 6.

\begin{center}
\psfig{figure=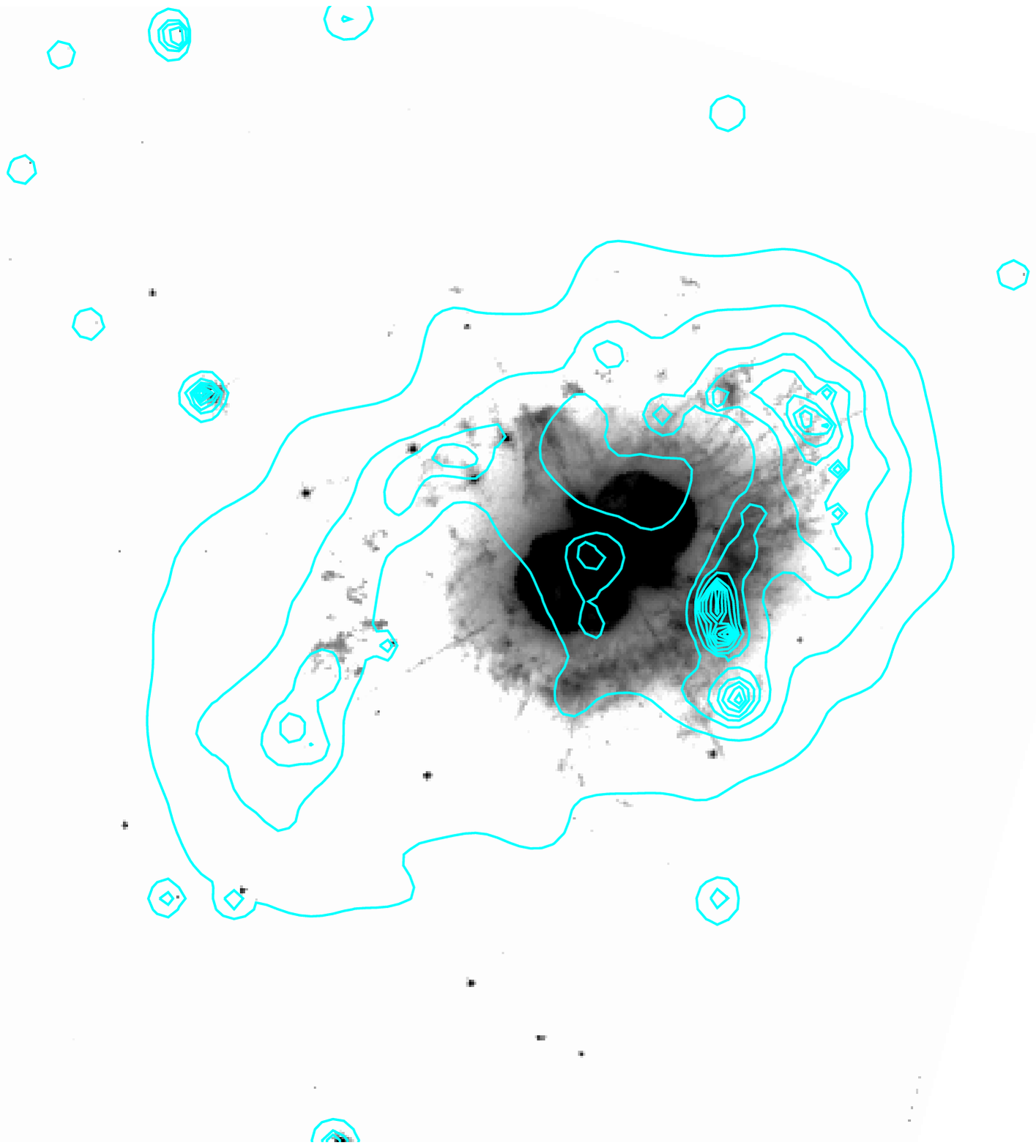,width=3.5in}
\begin{minipage}[h]{3.5in}
{\small
Figure 2: The soft X-ray ring overlaid on a 
6590 $\AA$ HST image of $\eta$
Car printed to show the Outer Shell.  The X-ray ring is clearly
associated with the ragged optical knots which form the Outer Shell.  
The brightest X-ray knot coincides with the 
brightest optical knot southwest of the central source.  Contour
spacing is proportional to the square root of the surface brightness. 
} 
\end{minipage}
\end{center}

\begin{center}
\psfig{figure=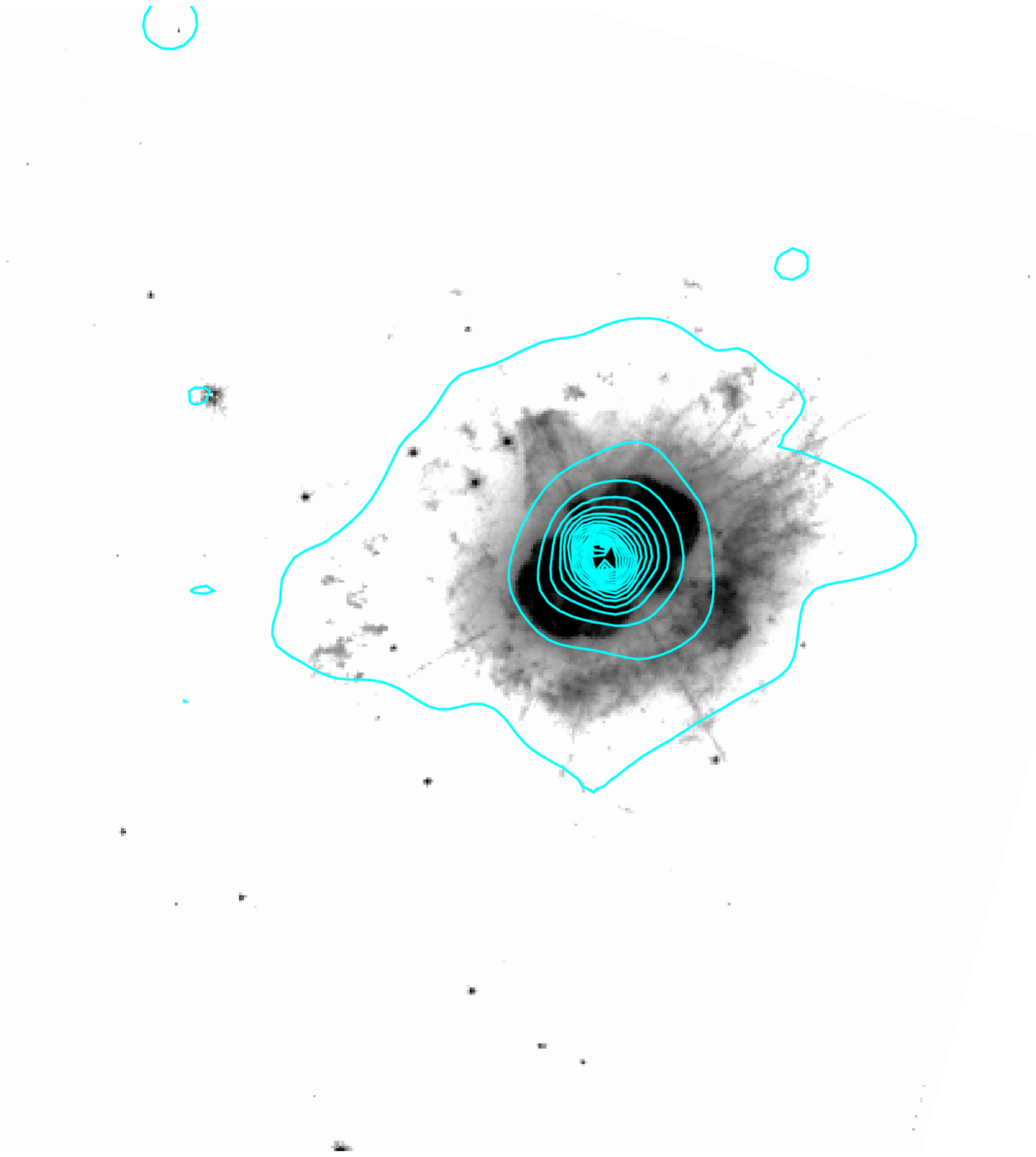,width=3.5in}
\begin{minipage}[h]{3.5in}
{\small
Figure 3: Contours of 2.1-8.0 keV surface brightness overlaid on a 
6590$\AA$ HST image of $\eta$ Car printed to show the Outer Shell.
Contour spacing is proportional to the square root of the surface 
brightness.  The bright X-ray source is coincident with the central
source of the bipolar nebula.  Closely-spaced contours around
this source indicate scattering from the telescope mirror but
the faintest contour partially follows the the Outer Shell. 
} 
\end{minipage}
\end{center}

\section{The Central Source}

The central source has a strength of 1.6 counts s$^{-1}$ in the CC mode
observation.  The TE-mode data, with apparent rate 0.17 s$^{-1}$, is
severely distorted by ``pileup'', a phenomenon where 2 or more photons
register in a single ACIS pixel during its recording interval and are
counted as one event having the summed energy of all interacting
photons in that pixel.  The flux measured at the detector 
is $4.4 \times 10^{-11}$ 
erg cm$^{-2}$ s$^{-1}$ and, at 2.3 kpc distance, the source luminosity
(before absorption in surrounding material and ISM) is L$_X$ = 
$6 \times 10^{34}$ erg s$^{-1}$ in the range 0.2-10 keV.  Flux
calculations using the PIMMS program (CXC et al,2000) are given in Table 1.  \\

Figure 4 shows the pulse-height spectrum of the central source.
Since there is no exact calibration, the energy scale is still 
uncertain.  If the Ir edge at 2.1 keV is used to determine the energy,
the energy of the prominent Fe-line emission is in the range 6.5 to
6.8 keV.  With this normalization, the general shape of the spectrum
is well reproduced using either a power law or a
thermal continuum.  There are emission lines at energies 2.4 (S),
3.7-4.0 (Ca),
and 6.5-6.8 (Fe) keV.  If the calibration assumed reproduces the
detector response reasonably, there may also be lines at 1.3 (Mg),
1.9 (Si), and 3.2 (A)
keV.  An absorption of $4\times 10^{22}$ atoms cm$^{-2}$
of cold material is required.  The continuum fit is not sensitive to 
temperature in the range kT = 6-20 keV.  If a
Raymond model is used, the Fe emission line shape seems to require a
strong contribution from Fe XXV and/or Fe XXVI at $\approx $ 6.8 keV and kT of
$\sim$ 15 keV, although both XTE and ASCA spectra indicate a temperature of
$\sim$ 5 keV (Corcoran et al, 2000a). \\

\begin{center}
\psfig{figure=goodfit_brems.ps,angle=-90,width=3.5in}
\begin{minipage}[h]{3.5in}
{\small
Figure 4: X-ray spectrum of the central source taken 
from the CC-mode data.  Spectrum is fit with
a Bremsstrahlung continuum with kT=15 keV.  The notch at channel 140
(2.1 keV)  is an absorption edge in the Ir mirror coating.  There is
a broad Fe emission line at channel 440 ($\approx $6.7 keV) and
evidence for emission lines at lower energies.  There is no emission
below 1.0 keV. 
} 
\end{minipage}
\end{center}

The broad Fe line emission requires at least two components at $\sim$
6.8 and $\sim$ 6.5 keV.  This is consistent with the report of Corcoran
et al (1998) of an Fe
fluorescence line at $\sim$ 6.4 keV in the ASCA data.  
A deeper Chandra observation is planned with the grating
and a well-calibrated ACIS configuration so we will not elaborate here on
the detail of these line emissions. \\

The appearance of the central source in the TE data (Figure 1)
suggests a surrounding halo of scattered X-rays. 
Figure 5 shows the Chandra point-response function (prf)
compared to the radial surface brightness     
of the central source.  The prf has been weighted according to the
measured spectrum and normalized to the counting rate measured in
the CC observation.  Since pileup is most severe in pixels with highest
counting rate, the central surface brightness is greatly
diminished, but there is no distortion apparent at radial
distances $>2.5^{\prime \prime}$.  The surface brightness follows the
prf closely from 2.5 to $5^{\prime\prime}$ radius, so we conclude that
there is no observable scattering halo at these radial distances.
Even excluding energies below 1.0 keV, some emission remains at
larger radial distances.  The faintest contour in Figure 3           
follows the morphology of the Outer Shell indicating some
emission from material in this area. \\

\begin{center}
\psfig{figure=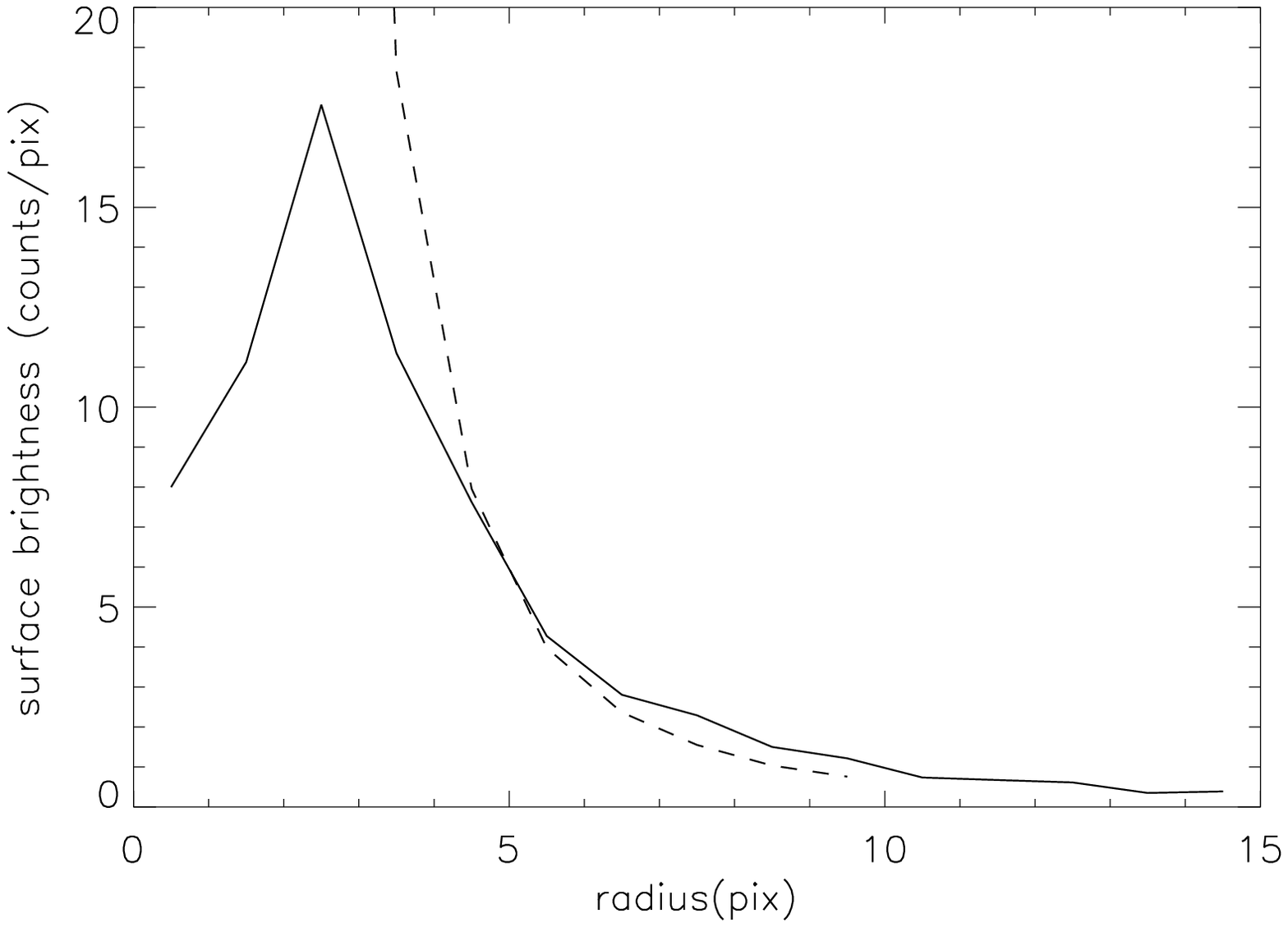,width=3.5in}
\begin{minipage}[h]{3.5in}
{\small
Figure 5: The radial surface brightness of the central source
compared to a weighted Chandra point-source response function.  The
solid line is the data, greatly depressed at the center because of
pileup.  Dashed curve is the prf which has intensity of 10,000 at the
source position.  One ACIS pixel = 0.5''. 
} 
\end{minipage}
\end{center}

The line of sight to the central source passes through the SE
(nearer) lobe of the bipolar nebula (Morse et al., 1998).  The path
length inside this lobe is $\approx 10^{17}$ cm so the observed
X-ray-absorbing column of $4 \times 10^{22}$ atoms cm$^{-2}$ implies a
density of $\approx 4 \times 10^{5}$ atoms cm$^{-3}$ inside the lobe.
This is in
good agreement with other estimates (Davidson and Humphreys, 1997).  As
the lobe expands, and if no more material is injected, the X-ray
absorbing column should decrease with time.

\section{The Outer Shell}

The emission is soft.  At energies above 2 keV the Shell almost disappears
(note the faint residual seen in Figure 3).  
The central source obscures any high-energy diffuse emission
which might come from the central region. \\

The geometry is more ring-like than shell-like.  The projection of a
limb-brightened shell would have more central emission.  A shell with
thickness 1/5 the radius would have central brightness $\approx$ 40\%
of the brightest part of the limb.  We observe only $\approx$ 15\%.
For the analysis, we
assume a ring (torus) with diameter 45$^{\prime\prime}$ = 0.50 pc and
thickness 
(diameter of circular cross-section) $7^{\prime\prime}$ = 0.08 pc.
The ring is inclined 45$^{\circ}$-50$^{\circ}$ from the plane of the
sky and 
1/4 of the ring is missing.  Volume of this partial ring is $1.7\times
10^{53}$ cm$^{3}$.  The brightest knot of emission has projected dimension
$4^{\prime\prime}\times 6^{\prime\prime}$ and, assuming an
intermediate
thickness, the volume is $2.6\times 10^{51}$ cm$^{3}$. \\

The spectrum of the Shell was obtained by extracting events from an
elliptical region including the Shell and excluding events from the
central source.  This spectrum shown in Figure 6 is a rather
smooth continuum with some structure from 0.6 to 0.9 keV.  This can be
fit with a bremsstrahlung continuum with kT ranging from 0.2 to 0.4
keV and with lines at $\sim 0.4$ and $\sim 0.7$ keV, perhaps from
N and Fe L.  A Raymond model
with cosmic abundances predicts line emission from Ne at $\sim 0.9$
and from Mg at $\sim 1.3$ which are
not evident in the spectrum.  We are not being precise here because the
calibration uncertainty increases at low energies.  Both effective
area and energy scale are not well-calibrated so the existence and
identification of spectral features is highly uncertain.
Nevertheless, a low energy feature in
ASCA spectra was also found by Tsuboi et al (1997) and by Corcoran et
al (1998) which they identified as due to N.  If this is really a N
emission line, the low transmission of the ISM at $\sim 0.4$ keV
implies that the Shell is remarkably bright in N emission, an appealing
thought since some of the optical knots shine brightly with N line
emission (Davidson et al, 1982; Davidson et al,,1986). 
The X-ray spectrum of the bright knot is, within the statistics of the $\approx
400$ counts collected, the same as that of the whole Shell. \\ 

\begin{center}
\psfig{figure=shell_spec.ps,angle=-90,width=3.5in}
\begin{minipage}[h]{3.5in}
{\small
Figure 6: X-ray spectrum of the ring.  Spectrum is fit with
a Bremsstrahlung continuum with kT=0.21 keV and with 2 emission lines
with energies $\approx$ 0.3 and 0.7 keV. 
} 
\end{minipage}
\end{center}

The X-ray flux from the Shell is $2.2\times 10^{-12}$ erg cm$^{-2}$
s$^{-1}$, measured at the detector and which, using kT = 0.3 and N$_{H}$ =
$2\times 10^{21}$ atoms cm$^{-2}$, implies an X-ray luminosity,
$L_{X}$ = $1.0 \times 10^{34}$ erg s$^{-1}$.  Assuming a uniform
density throughout the torus, we calculate a density of 50 atoms
cm$^{-3}$.  Total thermal energy is $2\times 10^{46}$ ergs.  Total
mass is $1.0\times 10^{-2}$ M$_{\odot}$. The bright knot contains
$4 \times10^{-4}$ M$_{\odot}$ at a density of 130 atoms cm$^{-3}$. \\

Davidson et al (1986) estimated the amount of UV-emitting material in
the S condensation as $\sim .02 M_{\odot}$ with electron density of 
$\sim 10^4$.  The X-ray emitting gas in the bright knot thus
represents only 1\% of the material in this region.

Even though the morphology is not convincingly shell-like, could this
material be a limb-brightened shell of hot gas behind a blast wave
originating 
in the Great Eruption 160 years ago?  The answer is yes.
The age of a blast wave  can be calculated from the radius and
temperature (e.g. Seward \& Velusamy, 1995) and the result is 
$185^{+105}_{-70}$ years which is consistant with the time since
the Great Eruption.

An alternate way of heating the X-ray emitting gas is via
the optical knots in the Debris Field.  These have velocities of
300-1300 km s$^{-1}$ (Walborn, Blanco, and Thackeray 1978) 
implying ejection (with no deceleration) 130-900 years ago.  The hot gas
in the Shell could be pre-existing material heated by energy 
deposited by these fast-moving knots, some of which may have
originated in the Great Eruption.  Gas temperature and knot velocity
are consistant with this process.  Indeed, all the knots may have
been ejected in 1843 (or even later) with high velocities but have 
since been decelerated in the Outer Shell, some more than others.  

The optical knots show strong line
emission which is thought to be due to shocks -- either from
high velocity knots colliding with interstellar clouds or a fast wind
impacting on cometary knots of material.  The X-ray emitting gas is
compatable with either process. \\

\section{Application to Model of The Central Source}

This Chandra observation has confirmed, in a satisfactory way, the
two-component model used to explain all observations since that of
Einstein; two components, soft and hard, with only the harder radiation
coming from the near vicinity of the central object.  We now compare
the 2-10 keV Chandra data with that of XTE which has been used to
define a promising model of the central source. \\

A square degree of sky containing $\eta$ Car has been monitored since
May 1997 with XTE (Corcoran, et al, 2000b).  The 2-10 keV flux observed
varies, usually smoothly, from 5 to $16\times 10^{-11}$ erg/cm$^{-2}$
s$^{-1}$ 
with some fluctuations at high intensity.  There is also a 3-month 
minimum, interpreted as an eclipse,
during which the flux drops, not to zero, but to a minimum of $0.5\times
10^{-11}$ erg/cm$^{-2}$ s$^{-1}$.  ASCA spectra obtained from a $3^{\prime}$
diameter region during this time yield a flux of $5.7\times 10^{-11}$
erg/cm$^{-2}$ s$^{-1}$ (2-10 keV) in 
the ``high'' state and $0.43\times 10^{-11}$ erg/cm$^{2}$ in the low,
eclipsed state (Corcoran, et al., 2000a).  If this 2-10 keV emission
observed during the eclipse is from $\eta$ Car itself, it has a
significant impact on the model. \\

The currently-proposed model ascribes the 2-10 keV X-ray flux to
shocked emission from a wind-wind collision in a massive binary
system.  In this binary model (Corcoran et al.  2000b), at the time of
the Chandra observation, the wind-wind collision shock is about 15 AU
$\approx 7\times10^{-3}$ arcsec from the more massive star, and so
would be unresolved to Chandra.  As noted by Corcoran et al. (2000a)
the ASCA X-ray spectrum of $\eta$ Car during the X-ray low state in
1997.9 is \textit{inconsistent} with the colliding wind binary model
since the colliding wind source should be completely hidden during the
eclipse (which is caused by absorption in the wind from the primary
star).  Perhaps, the 2-10 
keV emission seen by ASCA at that time is produced by another source
of hard emission in the ASCA extraction region (a circle of $3'$
radius around $\eta$ Car). We can search for this in the Chandra data.\\

The 2-10 keV flux measured by Chandra in CC mode (Fig. 4) is 
$4.4\times 10^{-11}$
erg/cm$^{-2}$ s$^{-1}$ and is almost all from the vicinity of the 
central source.  A
near-simultaneous observation with XTE (Corcoran, et al 2000) measures
$6\times 10^{-11}$ erg/cm$^{-2}$ s$^{-1}$.  
This discrepancy probably illustrates the
uncertainty in absolute flux measured by different instruments.  No
other significant sources of 
2-10 keV emission are seen in the field.  There are, however, at least
50 weak serendipitous sources (almost all are stars in the
cluster Tr16) 
within 5$^{\prime}$ of $\eta$ Car.  There is also weak diffuse
emission above 2 keV observed from the Outer Shell which is well
within the region covered by ASCA. \\

In the range 2-10 keV, we observe a small bright halo around the central
source which is due to the telescope point response function.  
Assuming that the flux from $\eta$ Car does not vary between
the CC and TE mode observations, the 2-10 keV diffuse flux 
at radial distances greater than 10$^{\prime\prime}$ is in excess of
that expected in the ``wing''
of the telescope point response function. \\

In an annulus defined by radii $10^{\prime \prime}$ and $90^{\prime
\prime}$, in the 2-10 keV band, the observed count rate is 5.5\% that
of the central source.  The Chandra mirror PRF indicates that mirror
scattering should be $\approx3.5$\%.  The 9 brightest serendipitous
sources account for 0.2\% so there is a residual diffuse
emission intrinsic to the region with strength $\approx$ 2\% that of
the central source or about $8\times 10^{-13}$ erg cm$^{-2}$
s$^{-1}$.  The ASCA-measured residual during eclipse minimum was
$\approx$ 5\% of the maximum source strength.  We conclude that,
although half of the ASCA-measured eclipse residual might be diffuse, 
the other half must come from
a region very close to $\eta$ Car, within a Chandra resolution
element of $\sim 0.5^{\prime\prime}$ or within $\sim 1000$ AU of 
the central star.
Speckle interferometry from the ground (Weigelt and Ebersberger 1986)
and direct imaging from space (Morse et al.  1998) have revealed the
presence of starlike knots, the Weigelt Blobs,  within 0.4$''$
from $\eta$ Carinae; it is possible that scattering of the colliding
wind emission by one or more of these knots produces the 2-10 keV
emission seen in the ASCA low state spectrum, and may be also
responsible for the Fe fluorescent line detected in the ASCA
spectra.  It is also possible that the X-ray emitting region is not
completely eclipsed and the model needs to account for this.

Acknowledgments:  
We thank Roberta Humphreys for helpful
suggestions and for pointing out several vital references, thus saving
us the embarassment of seeming to ignore work which defines the nature of
this peculiar object.  Financial support was provided by the Chandra
X-ray Center NASA Contract NAS8-39073.


\begin{deluxetable}{lllll}
\footnotesize
\tablecaption{Count Rate, Flux, and Luminosity\label{tbl-1}}
\tablewidth{0pt}
\tablehead{
\colhead{} & \colhead{} & \colhead{Energy Range} 
  & \colhead{Central Source} & \colhead{Ring} \nl
}
\startdata
ACIS Rate&counts s$^{-1}$&0.2-10 keV&$1.6\pm 0.1$&$0.40\pm .01$ \\
kT&keV&-&$15^{+5}_{-10}$&$0.3\pm 0.1$ \\
N$_{H}$&atoms cm$^{-2}$&-&$4\times 10^{22}$&$2\times 10^{21}$ \\
Flux (absorbed)&erg cm$^{-2}$ s$^{-1}$&0.2-10&$4.4^{+0.0}_{-0.7}\times 10^{-11}$&$2.2\pm 0.3\times 10^{-12}$ \\
Flux (unabsorbed)&erg cm$^{-2}$ s$^{-1}$&0.2-10&$8.8\pm 0.3\times 10^{-11}$&$1.5^{+1.7}_{-0.5}\times 10^{-11}$ \\
Flux (absorbed)&erg cm$^{-2}$ s$^{-1}$&2-10&$4.3^{+0.0}_{-0.7}\times 10^{-11}$&- \\ 
L$_{X}$ (No ISM absorption)&erg s$^{-1}$&0.2-10&$2.8^{+0.0}_{-0.5}\times 10^{34}$&$1.0^{+1.1}_{-0.6}\times10^{34}$ \\ 
L$_{X}$ (No  $\eta$ Car absorption)&erg s$^{-1}$&0.2-10&$5.6\pm 0.2\times 10^{34}$&- \\

\enddata
\end{deluxetable}

\end{document}